\documentclass[fleqn,10pt]{wlscirep}
\usepackage{graphicx}
\usepackage{cases}
\usepackage{amssymb,amsfonts,amsmath,bm}
\title{Turbulent chimeras in large semiconductor laser arrays}

\author[1,*]{J. Shena}
\author[1]{J. Hizanidis}
\author[2]{V. Kovanis}
\author[1,3,4]{G.~P. Tsironis}
\affil[1]{Crete Center for Quantum Complexity and Nanotechnology, Department of Physics, University of Crete, 71003 Heraklion, Greece}
\affil[2]{Department of Physics, School of Science and Technology, Nazarbayev University, 53 Kabanbay Batyr Ave, Astana, Republic of Kazakhstan}
\affil[3]{Institute of Electronic Structure and Laser,
Foundation for Research and Technology--Hellas, P.O. Box 1527, 71110 Heraklion, Greece.}
\affil[4]{National University of Science and Technology MISiS, Leninsky prosp. 4, Moscow, 119049, Russia}
\affil[*]{jshena@physics.uoc.gr}

\begin{abstract}
Semiconductor laser arrays have been investigated experimentally and theoretically from the viewpoint
of temporal and spatial coherence for the past forty years.
In this work, we are focusing on a rather novel complex collective behavior, namely chimera states, where synchronized clusters 
of emitters coexist with unsynchronized ones.
For the first time, we find such states exist in large diode arrays based on quantum well gain media with nearest-neighbor interactions.
The crucial parameters are the evanescent coupling strength and the relative optical frequency detuning between the emitters of the array.
By employing a recently proposed figure of merit for classifying chimera states, 
we provide quantitative and qualitative evidence for the observed dynamics. The corresponding chimeras are identified as \emph{turbulent} according to
the irregular temporal behavior of the classification measure.
\end{abstract}
\begin{document}

\flushbottom
\maketitle
%
%
\thispagestyle{empty}


\section*{Introduction}

Semiconductor lasers are enabling components in multiple  platform applications 
spanning optical communication networks to laser surgery and sensing.
Recent works include impressive advances in high-speed lasers with low power 
consumption, high-power vertical external cavity surface emitting lasers and high-speed beam steering with phased vertical cavity laser 
arrays. Significant advances have been made in nitride based lasers, record-high temperature operation quantum dot lasers,
and the field of nanolasers with ultralow volume and threshold is coming to technological maturity~\cite{Urbana}.

Of special importance for next generation applications such as laser radars, is the design of photonically 
integrated semiconductor laser arrays that consist of a very large number of properly coupled photonic emitters~\cite{HEC13}.
It is well known that phase locking of an array of diode lasers is a highly effective method in beam shaping
because it increases the output power and reduces the overall needed lasing threshold. Recent work on phase-locked
laser arrays through global antenna mutual coupling has employed custom made nano-lasers~\cite{MIT_2016}.
Moreover, reconfigurable semiconductor laser networks based on diffractive coupling using Talbot geometry have
been studied on commercially available vertical cavity diode lasers~\cite{Fischer_2015}.

In the present work, we are interested in the collective behavior
of a large array of semiconductor lasers with nearest-neighbor interactions.
The crucial parameters for the observed dynamics are the coupling strength and the relative optical
frequency detuning between the lasers,
which introduces realistic inhomogeneities into the system.
Our focus, in particular, is to identify the parameter regions where chimera states emerge
and subsequently characterize these states using suitable classification measures~\cite{kevrekidis}.

Chimera states were first reported for identical and symmetrically coupled phase oscillators
~\cite{KUR02a}. For over a decade now, a number of works has been dedicated to 
this phenomenon of coexisting synchronous and asynchronous oscillatory behavior
(see \cite{panaggio:2015} and references within). The latest developments in this field
involve their study in physical, higher-dimensional systems beyond phase oscillators, their experimental
verification~\cite{tinsley:2012,hagerstrom:2012,wickramasinghe:2013,martens:2013,
Rosin2014,schmidt:2014,Gambuzza2014,Kapitaniak2014,HAR16}, their robustness against system inhomogeneities~\cite{LAI12,YAO13,OME15,HIZ16a},
their existence in stochastic systems~\cite{SEM16},
and their manipulation through control techniques~\cite{SIE14,BIC15,ISE15,OME16}.

Coupled lasers have been extensively studied in terms of nonlinear
dynamics \cite{mandel1,OLI01,UCH01,DAH12,Fischer} and synchronization phenomena~\cite{LYT97,PEC14,ALS96,JUN16},
but works on chimera states in laser networks have appeared only recently. 
In~\cite{LAR13,LAR15} chimera states were reported both theoretically and experimentally
in a virtual space-time representation of a single laser system subject to long delayed
feedback. Furthermore, so-called ``small chimeras'' were numerically observed in
a network of four globally delay-coupled lasers in~\cite{BOE15,ROE16}, for both small and large delays.
Such chimeras exist for very small network
sizes and do not require nonlocal coupling in order to emerge.
In our study we use neither nonlocal, nor global coupling but simple nearest-neighbor interactions
which is physically plausible for lasers, e.~g., grown on a single chip. 
This coupling realization is less expensive computationally.  Moreover, it
revises the general belief that nonlocal coupling is essential for the existence of chimeras~\cite{HIZ16}.

We will show that the crucial parameter for the collective behavior in our system is the frequency
detuning between the coupled lasers. The effect of detuning has been examined before in
~\cite{Dutta} but with respect to in- and anti-phase synchronization.
Moreover, transitions from complete to partial synchronization (optical turbulence) were
explained, for a small array of three lasers.
Here, we address the emergence of the hybrid phenomenon of chimera states in a \emph{large} laser array
and provide a quantification of these patterns using newly developed classification measures~\cite{kevrekidis}.
%

\section*{Results}
Our system consists of an array of $M$ locally coupled semiconductor lasers. 
The evolution of the slowly varying complex amplitudes ${\mathcal E}_{i}$ of the electric fields and 
the corresponding population inversions $N_{i}$ are given by~\cite{Arecchi,Wieczorek}:
\begin{eqnarray} 
\frac{d{\mathcal E}_{i}}{dt}&=&(1-ia){\mathcal E}_{i}N_{i}+i\eta({\mathcal E}_{i+1}+{\mathcal E}_{i-1})+i\omega_{i}{\mathcal E}_{i} \nonumber \\
T\frac{dN_{i}}{dt}&=&(p-N_{i}-(1+2N_{i})|{\mathcal E}_{i}|^2), \quad i=1\dots M \label{eq1}
\end{eqnarray}
The amplitude-phase coupling is modeled by the linewidth enhancement factor $a=5$, $T=400$ is the ratio
of the carrier to the photon lifetime of the photons in the laser cavity.
The normalized angular frequency $\omega_{i}$ measures the optical frequency detuning of laser $i$ from a common reference. The diode 
lasers are pumped electrically
with the excess pump rate $p=0.5$.
These parameters represent typical experimental values from multiple experiments performed in the past 20 years using quantum well laser media~\cite{Simpson}.
The coupling strength $\eta$ is a control parameter
used to tune the dynamics of the system.
We have used open boundary conditions to account for the termination of the array in a
finite system.
By using polar coordinates ${\mathcal E}_{i}=E_{i}e^{i(\phi_{i}+\omega_{i}t)}$ and separating real from imaginary part, we get: 
\begin{eqnarray} 
\frac{dE_{i}}{dt}&=&E_{i}N_{i}-\eta[E_{i+1}\sin(\Delta\phi_{i+1}+\Delta\omega_{i+1}t)
+E_{i-1}\sin(\Delta\phi_{i-1}+\Delta\omega_{i-1}t)] \nonumber \\
\frac{d\phi_{i}}{dt}&=&-aN_{i}+\eta[\frac{E_{i+1}}{E_{i}}\cos(\Delta\phi_{i+1}+\Delta\omega_{i+1}t) 
+\frac{E_{i-1}}{E_{i}}\cos(\Delta\phi_{i-1}+\Delta\omega_{i-1}t)] \nonumber  \\
T\frac{dN_{i}}{dt}&=&p-N_{i}-(1+2N_{i}){E_{i}}^2 \label{eq2}
\end{eqnarray}
where $\Delta\phi_{i+1}=\phi_{i+1}-\phi_{i}$, $\Delta\phi_{i-1}=\phi_{i-1}-\phi_{i}$, $\Delta\omega_{i+1}=\omega_{i+1}-\omega_{i}$, $\Delta\omega_{i-1}=\omega_{i-1}-\omega_{i}$.
For the special case of \emph{two lasers} and in the absence of detuning, Eqs.~(\ref{eq2}) have the following fixed points:
\begin{equation} \label{eq3}
E_{1}=E_{2}=\sqrt{p},  N_{1}=N_{2}=0,  \phi_{2}-\phi_{1}=0  
\end{equation}
\begin{equation} \label{eq4}
E_{1}=E_{2}=\sqrt{p},  N_{1}=N_{2}=0,  \phi_{2}-\phi_{1}=\pi 
\end{equation}
To investigate the stability of these steady states we introduce small perturbations and linearize
Eqs.~(\ref{eq2}) about their steady-state values \cite{Winful_1988}. The Routh-Hurwitz criterion is
used to determine the parameter value regions in which
the steady-state solutions are stable.
After some calculations we find that the fixed point of Eq.~(\ref{eq4}) is stable under the condition:
\begin{equation} \label{eq5}
\eta<\frac{1+2p}{2aT}
\end{equation}
and the fixed point of Eq.~(\ref{eq3}) is stable for:
\begin{equation} \label{eq6}
\eta>\frac{ap}{1+2p}
\end{equation}

\begin{figure}[]
\centering
\includegraphics[width=.85\textwidth]{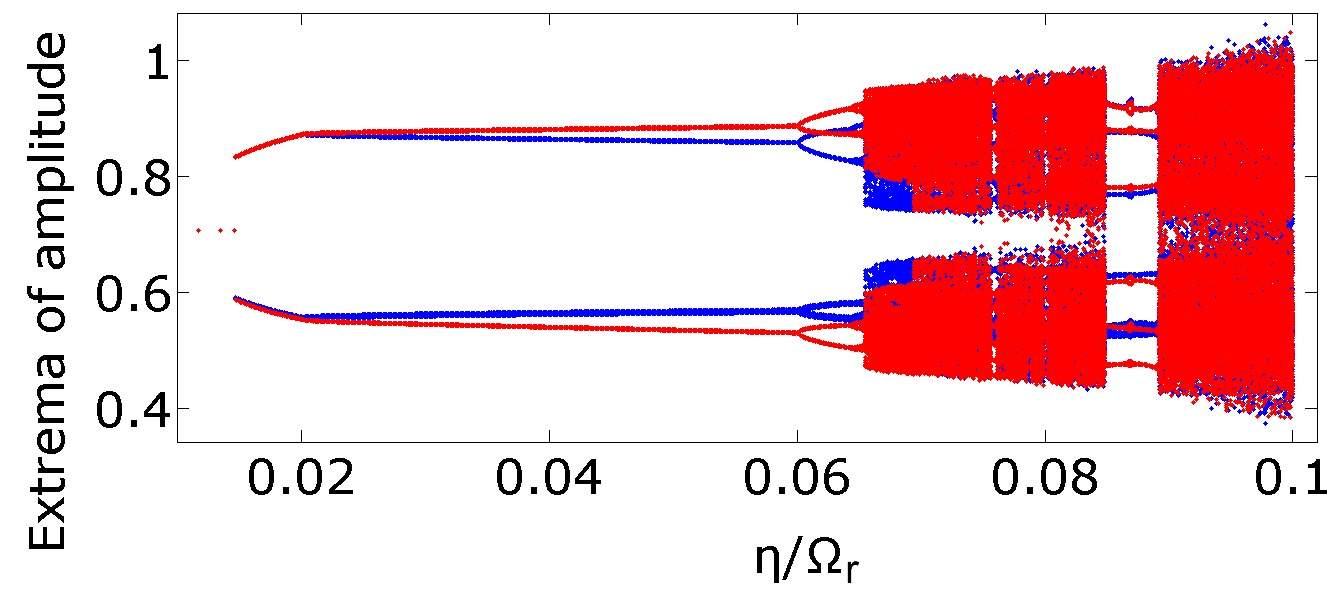}
\caption{This figure depicts the amplitude maxima and minima of the electric field of two coupled diode lasers 
in dependence of the coupling strength $\eta$ which has been rescaled
to the relaxation oscillation frequency $\Omega_r$. The blue color refers to the first laser
and the red color to the second one. The steady state, otherwise known as continuous wave operation, undergoes a Hopf bifurcation
at $\eta/\Omega_r=0.01$ and, as a result, a limit cycle is born, that oscillates at the free running relaxation frequency.
At ${\eta/\Omega_r=0.06}$, the system undergoes a period doubling bifurcation leading to a chaotic region
which is interrupted by windows of periodic operation. 
Other parameters are: $T\Omega_r=20$, $p=0.5$, $\frac{1}{\Omega_r}=20$, and $a=5$.\label{fig_1}}
\end{figure}
In order to understand the effect of coupling strength, Fig.~\ref{fig_1} depicts a numerically obtained bifurcation diagram of the maxima and minima of the amplitude of the oscillating electric field.
A Hopf bifurcation occurs at $\eta/\Omega_r=0.01$ where the coupling strength is normalized to the relaxation oscillation frequency $\Omega_r=\sqrt{2p/T}$~\cite{Erneux_1997}.  
As the coupling is increased the limit cycle exists until $\eta/\Omega_r=0.06$. After that, a period-doubling cascade takes place, leading to chaos. The system remains chaotic until the approximate
value of $0.084$ and then enters a new limit cycle which is stable up to $\eta/\Omega_r=0.089$,
which is followed by a new period doubling cascade into a second chaotic region.
\begin{figure}[]
\centering
\includegraphics[width=.85\textwidth]{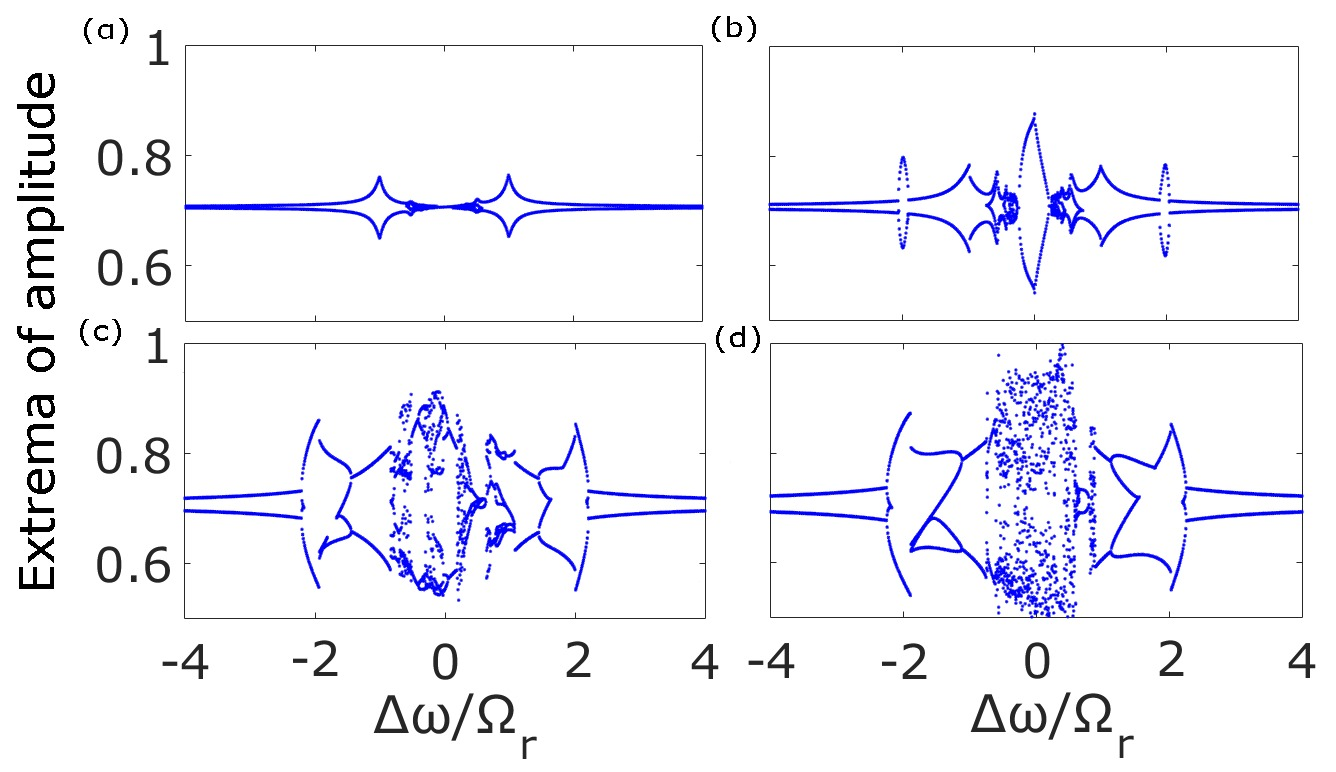}
\caption{Extrema of the amplitude of the electric field in dependence of the detuning, for different values of the coupling strength. (a) $\frac{\eta}{\Omega_r}=0.01, (b)\frac{\eta}{\Omega_r}=0.025, (c) \frac{\eta}{\Omega_r}=0.0635, (d) \frac{\eta}{\Omega_r}=0.08$. The difference in the detuning $\Delta\omega_r$ has been rescaled by the relaxation oscillation frequency $\Omega_r$. 
In (a) we notice that the amplitude resonates at ${\eta/\Omega_r=1}$, as expected. Such behavior shows the primary resonance of the system
when the parametric driving frequency is equal to the internal frequency of the oscillator, i.~e. the free running relaxation oscillation.
In (b) we notice subharmonic resonances at ${\eta/\Omega_r=\pm 2}$, hysteretic behavior at the primary resonance, and at (c) and (d) 
the emergence of chaos is evident.
Other parameters: $T\Omega_r=20, p=0.5, \frac{1}{\Omega_r}=20$ and $a=5$.
\label{fig_2}}
\end{figure}

Apart from the coupling strength, another crucial parameter is the optical frequency detuning and its correlation with the amplitude instability and mutual coherence of the light emitted by the laser.
For both solid state~\cite{Roy_1997,Roy_2007} and semiconductor lasers~\cite{Dutta},
the complexity of the system increases immensely by introducing detuning.
As expected, the most relevant parameter is actually the \emph{difference} between the laser detunings rather than their individual values. The bifurcation diagram of Fig.~\ref{fig_2} shows the maxima and minima of the 
electric field amplitude in dependence of $\Delta \omega=\omega_{2}-\omega_{1}$, rescaled by the free relaxation frequency $\Omega_r$.
This has been repeated for various values of the coupling strength (Figs.~\ref{fig_2}(a-d)).
We observe that in a certain range of $\Delta \omega/\Omega_r$ values
the amplitude of the laser oscillations increases significantly. Moreover, for large coupling strengths (Figs.~\ref{fig_2}(c,d))
the behavior of the system is rich and complex in dynamical responses. 
It is also noticeable that although some $\eta$ values render the system chaotic
in the case without detuning (see Fig.~\ref{fig_1}), for the same coupling strengths 
the dynamics is regular in the presence of detuning (Figs.~\ref{fig_2}(d)).

The situation is much more complicated when we consider larger arrays.
In the case of $M$ coupled lasers, it can be found that the critical coupling strength,
for the special case of the anti-phase region (see Eq.~(\ref{eq6}) for two coupled lasers) changes to~\cite{Winful_1992}:
\begin{equation} \label{eq7}
\eta<\frac{1+2p}{4aT\cos\left(\frac{\pi}{M+1}\right)}
\end{equation}
As $M$ increases, the critical coupling decreases roughly as $\frac{M}{M-1}$ and
reaches a limiting value at large $M>10$ which is half of that corresponding to $M=2$.
Throughout this work, we will consider an array of $200$ lasers. The numerical integration has been done by using the fourth order Runge-Kutta algorithm.

\subsection*{Dynamics of coupled lasers with zero detuning}
\begin{figure}[]
\centering
\includegraphics[width=.85\textwidth]{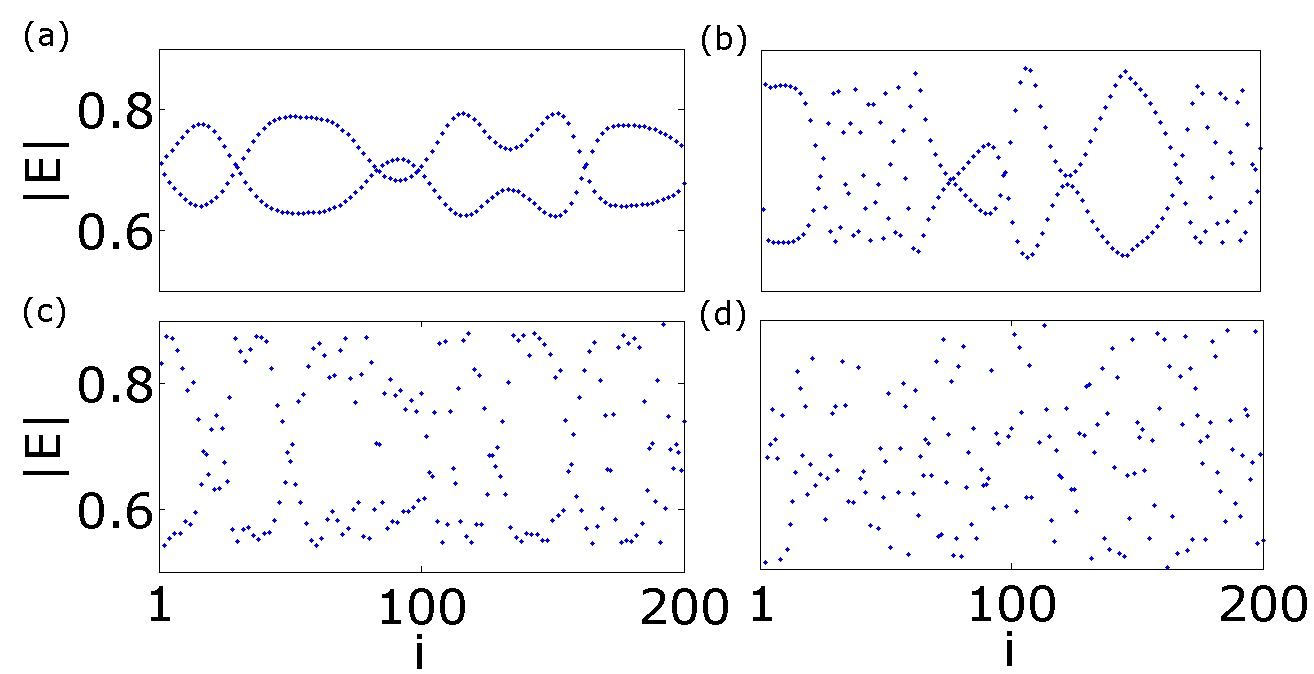}
\caption{Snapshots of the amplitude of the electric field in an array of $M=200$ lasers for different coupling strengths without detuning:
$(a) \frac{\eta}{\Omega_r}=0.006,  (b) \frac{\eta}{\Omega_r}=0.01, (c) \frac{\eta}{\Omega_r}=0.02. (d) \frac{\eta}{\Omega_r}=0.07$. 
Other parameters: $T\Omega_r=20, p=0.5, \frac{1}{\Omega_r}=20$ and $a=5$. For low coupling strengths each laser is in
anti-phase synchronization with its nearest neighbors (panel (a)). As the coupling increases the system enters the fully incoherent state
(panel (d)).
\label{fig_3}}
\end{figure}

First we focus on the influence of the coupling strength on the collective behavior, in the 
absence of detuning. We use the same initial conditions throughout the manuscript, namely random phases 
taken from a uniform distribution on the interval [$-\pi$ to $\pi$], and fixed amplitudes $E_i=\sqrt{p}$ and inverse populations $N_i=0$.
According to Eq.~(\ref{eq7}), the Hopf bifurcation for our laser array occurs at the value $\eta/\Omega_r=0.005$.
Slightly above this value, the system demonstrates a self-organized pattern (see Fig.\ref{fig_3} (a-b)):
The laser array splits into two sub-systems with each laser
having a phase difference equal to $\pi$ with its nearest neighbors (anti-phase synchronization~\cite{Dutta}).
This pattern gradually vanishes with increasing coupling strength and the system becomes fully incoherent (Figs.~\ref{fig_3}(c-d)). 
In Fig.~\ref{fig_3} snapshots of the amplitude of the electric field are shown at $100T_r$, where
$T_r=2\pi/\Omega_r$ is the period of the relaxation oscillation of the free running diode laser.
\subsection*{Effect of optical frequency detuning and chimera states}
The situation becomes significantly different when we consider finite optical frequency detuning.
We incorporate detuning in the following way:
\begin{equation} \label{eq8}
\frac{\omega_{i}}{\Omega_r}=\Delta i
\end{equation}
where $\Delta$ is a constant.
\begin{figure}[]
\centering
\includegraphics[width=.85\textwidth]{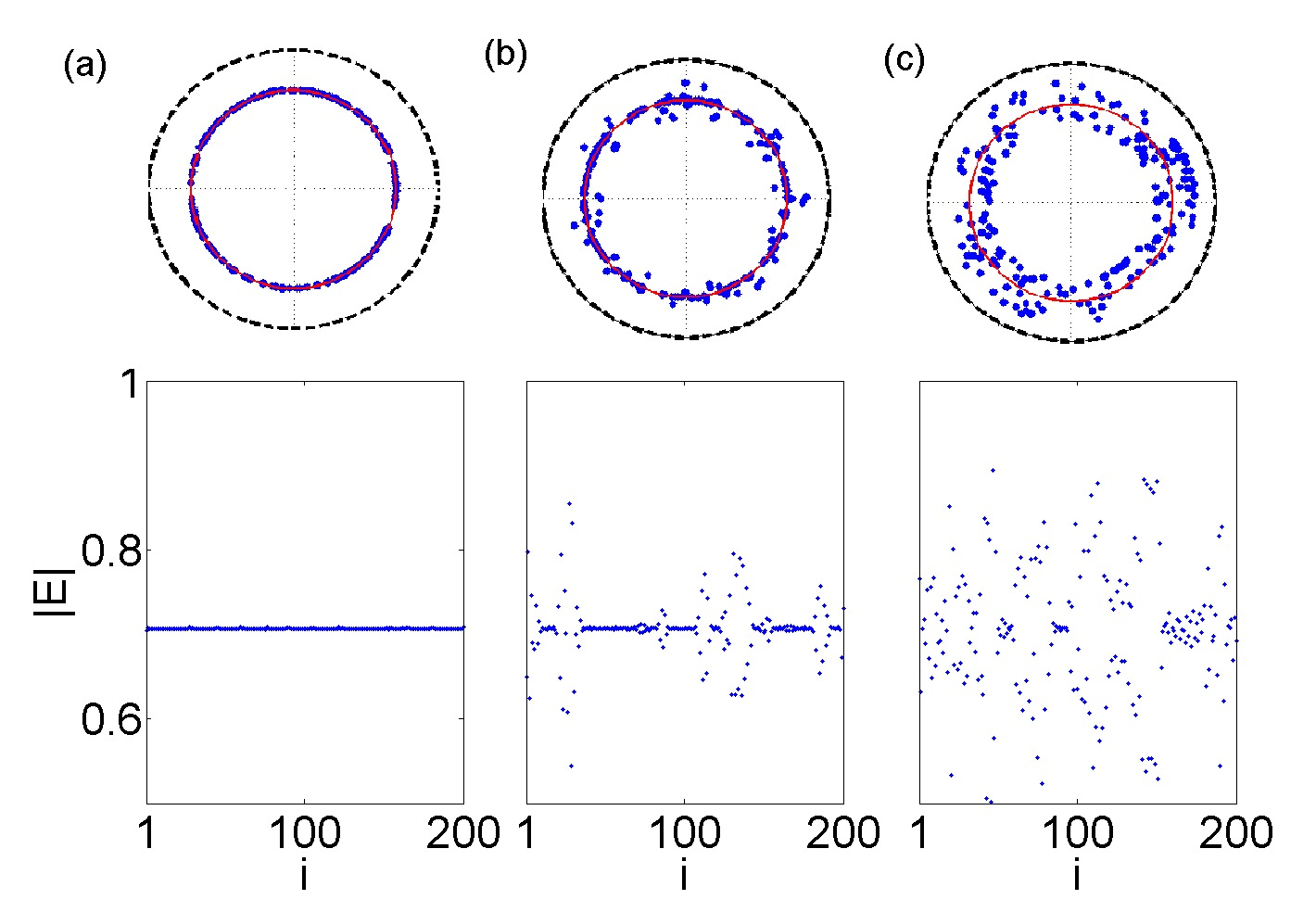}
\caption{Top: The electric field
in the complex unit circle for different coupling strengths and constant detuning. The red circle denotes the steady state solution where the amplitude of the oscillations is constant. Bottom: Corresponding snapshots of the amplitude of the electric field. (a) $H=0.008$ (fully synchronized state), (b) $H=0.014$, (amplitude chimera state), and (c) $H=0.026$ (incoherent state).  Other parameters: $\Delta=0.01, T\Omega_r=20, p=0.5, \frac{1}{\Omega_r}=20$ and $a=5$.
For further visualization refer to the Supplementary Movies S1-S3. \label{fig_4}}
\end{figure}
With this distribution, the differences of the detuning have a simple form:
$\left |\Delta\omega_{i+1}/\Omega_r\right |=\left |\Delta\omega_{i-1}/\Omega_r\right |=\Delta$
~\cite{OLI01}. 
It is possible to realize different forms
of synchronization depending on the coupling strength, which we redefine as $H=\frac{\eta}{\Omega_r}$. One case is full synchronization,
where $E_{i} = E_{j}$ holds
for all lasers $i, j = 1 \dots M$ (see Fig.\ref{fig_4}(a), bottom). The behavior is therefore similar to that
of the uncoupled system since the whole array ends up in the steady state (each laser is lasing with
constant intensity equal to $\sqrt p\sim0.7$). 
In a partially synchronized state the amplitudes are different in one or more lasers (see
Fig.~\ref{fig_4}(b), bottom) and in the unsynchronized state there
is no fixed amplitude relation between the oscillators (see Fig.~\ref{fig_4}(c), bottom).
In Figs.~\ref{fig_4}(a-c) (top) we can see all of these states depicted in the complex unit circle.
The red circle denotes the steady state solution where the amplitude of the oscillations is constant. 
In the top panel of Fig.~\ref{fig_4}(a) the amplitudes are locked to this value, while the phases of the individual lasers are randomly 
distributed over the steady state solution circle. This case corresponds to amplitude (intensity) synchronization.
The opposite situation is full asynchrony, displayed in the top panel of Fig.~\ref{fig_4}(c), where both amplitude and phase 
exhibit incoherent behavior. The intermediate case is shown in Fig.~\ref{fig_4}(b) where an amplitude-chimera~\cite{ZAK14}
is illustrated through the coexistence of partial amplitude locking and incoherence. 
(For more information see Supplementary Movies S1-S3 corresponding to Figs.~\ref{fig_4}(a-c)).
\begin{figure*}[]
\includegraphics[width=\textwidth]{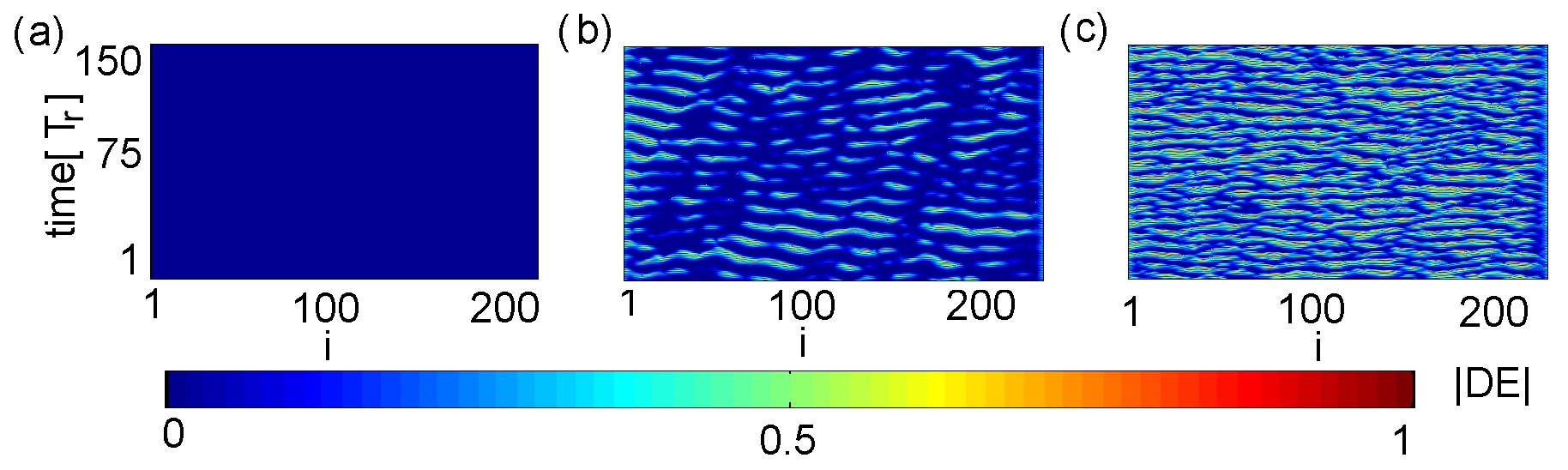}
\caption{Spatio-temporal evolution of the local curvature $DE_{i}$ (Eq.~(\ref{eq9})) for different values of the coupling strength: (a) $H=0.008$, (b) $H=0.014$, (c) $H=0.026$. Blue and red color denote full synchronization ($DE_{i}=0$) and full incoherence ($DE_{i}=0$), respectively. The spatio-temporal representation of the turbulent chimera state is shown in the middle panel.  
Other parameters as in Fig.~\ref{fig_4}. \label{fig_5}}
\end{figure*}

In order to quantify the spatial coherence of the observed patters we calculate the
local curvature $DE_{i}$ ( Eq.~(\ref{eq9})). Figure~\ref{fig_5} shows the spatio-temporal evolution of the local curvature corresponding to the states of Fig.~\ref{fig_4}.
In the fully synchronized case the local curvature is equal to zero (Fig.~\ref{fig_5}(a)). In Fig.~\ref{fig_5}(b) 
we have the case of an amplitude-chimera state. We see that this is not a stationary pattern since the local curvature oscillates in time. 
The fully incoherent states is shown in Fig.~\ref{fig_5}(c), where the local curvature attains higher values.

\begin{figure}[]
\centering
\includegraphics[width=.85\textwidth]{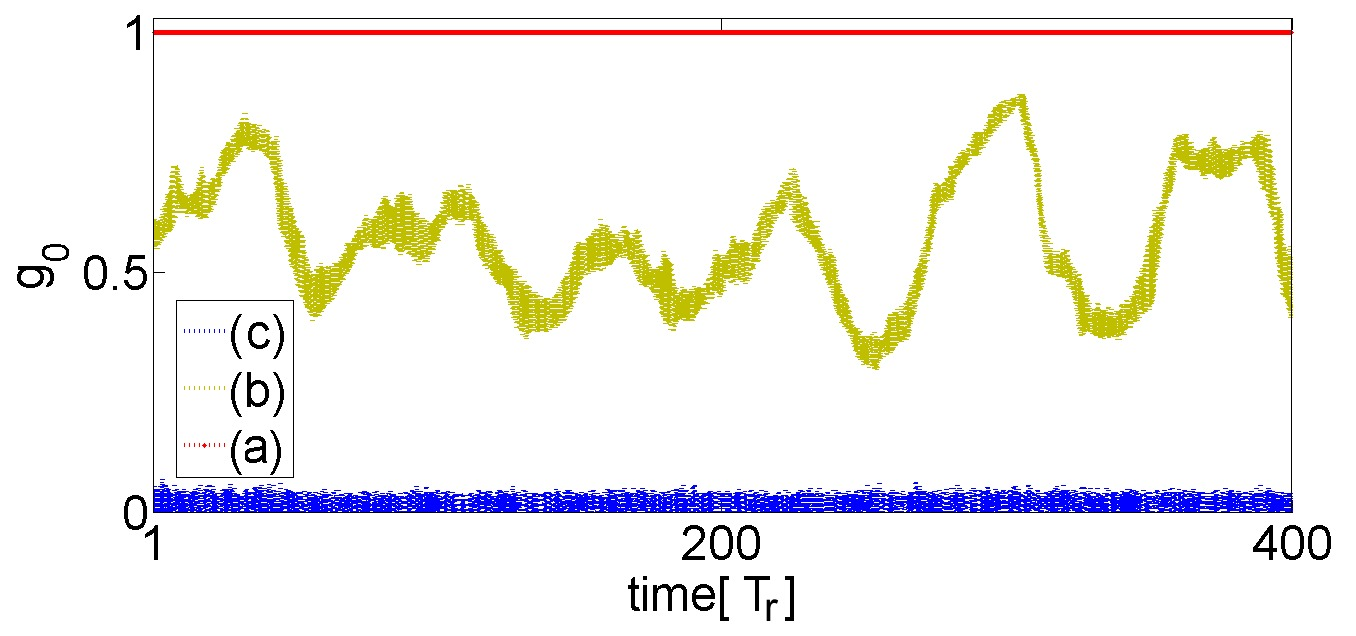}    
\caption{The time evolution $g_{0}(t)$  (Eq.~\ref{eq:g_0}) of the spatial extent occupied by the coherent lasers, for Figs.~\ref{fig_5}(a-c). 
In the fully synchronized state $g_{0}(t)$ is constant and equal to unity (a). The irregular oscillatory $g_{0}(t)$ is a signature for
a turbulent chimera state (b). The incoherent state corresponds to $g_{0}(t)$ close to zero (c). Other parameters as in Fig.~\ref{fig_4}. \label{fig_6}}
\end{figure}

In Fig.~\ref{fig_6}, the time evolution of the spatial extent occupied by the coherent lasers, $g_0(t)$ (Eq.~\ref{eq:g_0}),
for all three cases of Fig.~\ref{fig_5} is plotted. We see that for the case of Fig.~\ref{fig_5}(b) $g_0$
oscillates in an irregular manner, and therefore the corresponding amplitude chimera states are \emph{turbulent}
according to the classification scheme in~\cite{kevrekidis}. The other two curves (a) and (c) refer to full synchronization and full incoherence, respectively.
\begin{figure*}[]
\includegraphics[width=.85\textwidth]{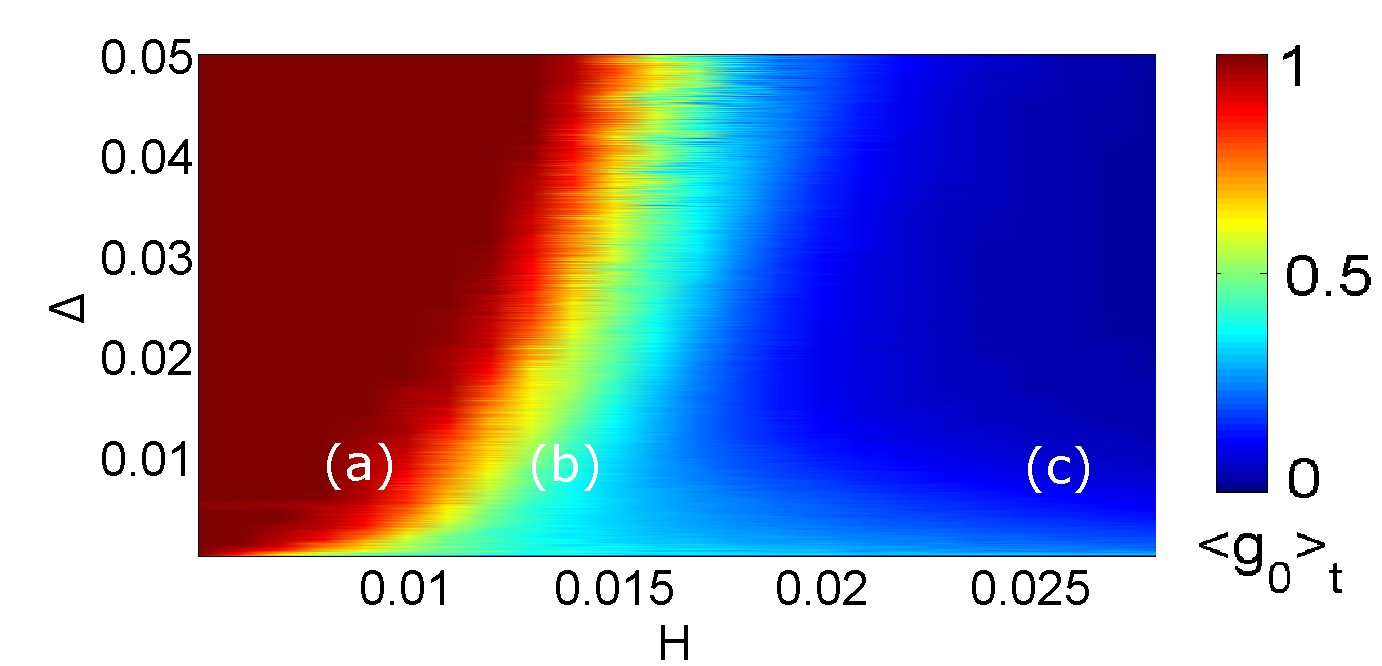}
\caption{Dependence of the temporal mean ${\langle g_{0} \rangle}_{t}$ on parameters $H$ and $\Delta$. Points (a) $(H=0.008,\Delta=0.01)$, (b) $(H=0.014,\Delta=0.01)$, and (c) $(H=0.026,\Delta=0.01)$, correspond to Figs.~\ref{fig_5}(a-c). The boundary between full synchronization (red) and 
full desynchronization (blue) marks the regions where turbulent chimeras emerge. Other parameters as in Fig.~\ref{fig_4}.\label{fig_7}}
\end{figure*}

The coaction of the detuning and the coupling strength on the observed 
synchronization patterns will be discussed next. In Fig.~\ref{fig_7} the temporal mean of $g_0(t)$ (averaged over $400T_r$) is plotted
in the ($H,\Delta$) parameter space. The initial conditions of the phases are randomly distributed between $-\pi$ and $\pi$, while for the electric field amplitudes and the population inversions they are chosen identical for all lasers: $E_{i} = \sqrt{0.5}$, $N_{i} = 0$. 
The labels (a), (b) and (c) mark the coordinates corresponding to Figs.~\ref{fig_5}(a),~\ref{fig_5}(b) and~\ref{fig_5}(c), respectively.
It is clear, that the parameter space is separated in two main domains, one of $<g_0>_t$ values 
close to unity which corresponds to full coherence and contains point (a), and one of $<g_0>_t$ values tending to zero which corresponds 
to full incoherence and contains point (c).
On the boundary between these two areas, lies a small region where the amplitude chimeras arise.
Note that, due to multistability, the mapping of the dynamical patterns may slightly change with different choice of initial conditions.
The qualitative result, however, will be the same.
\begin{figure}[]
\centering
\includegraphics[width=\textwidth]{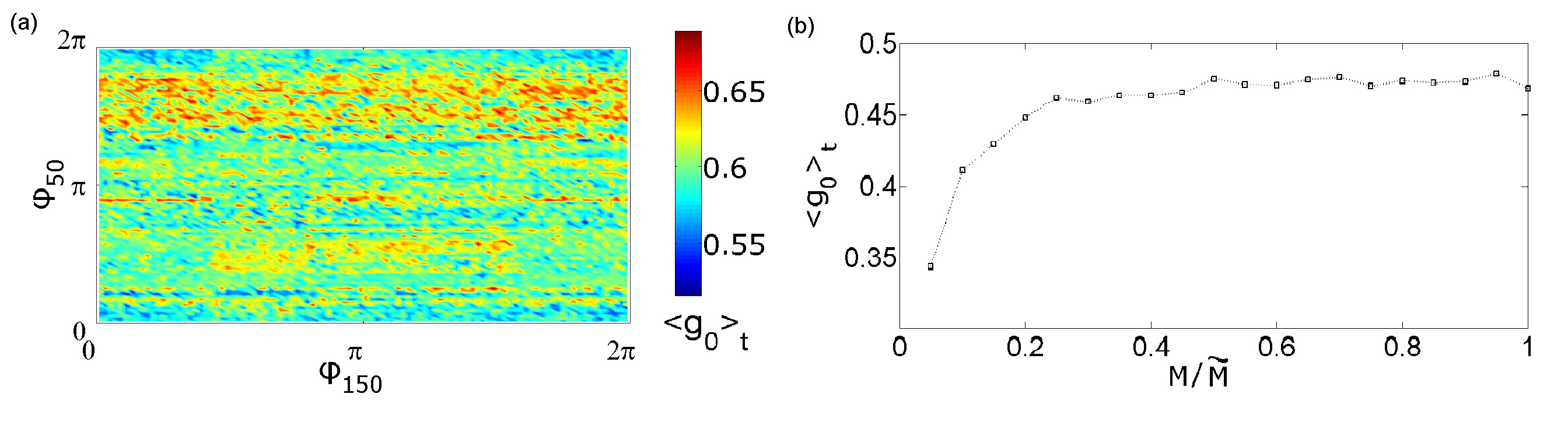}
\caption{(a) The temporal mean $<g_{0}>_{t}$ on the ($\phi_{50},\phi_{150}$)-projection. The initial conditions of the phases
for all lasers are random and fixed except for $\phi_{50}$ and $\phi_{150}$ which are varied.  
(b) The temporal mean $<g_{0}>_{t}$ as a function of the system size normalized to $\widetilde M=1000$. 
Parameters: $H=0.014, \Delta=0.01, T\Omega_r=20, p=0.5, \frac{1}{\Omega_r}=20$ and $a=5$.\label{fig_8}}
\end{figure}
For example, in Fig.\ref{fig_8} (a), we plot $<g_0>_t$ for a system with all initial phases randomly distributed but fixed, 
except those of laser 50 and 150, which we vary from $0$ to $2\pi$.
Clearly, the exact values of $<g_0>_t$ change but remain within the range allowing for chimera states.   

Finally, the question of system size is addressed.
In our simulations we observe that the behavior of the system does not change significantly when increasing $M$ from $200$ to $1000$. This is
illustrated in Fig.\ref{fig_8} (b). After $M>200$ the temporal mean $<g_{0}>_{t}$ remains constant in time.
From this fact we can conclude that, for an appropriately large system, the formation of chimera states is size-independent.

\section*{Discussion}
\label{discussion}
In conclusion, we have found amplitude chimera states in a large one-dimensional network
of semiconductor lasers by properly modifying the optical frequency detuning. Local coupling is sufficient to generate these states.
By using suitable classification measures we have quantified the observed dynamics.
Due to the system's multistability, even a slight change in the initial conditions
may produce different values for these measures. However, the range of the obtained values
ensures the existence of chimeras, the nature of which is turbulent.
The system size also has an effect on the calculated values, which saturate for 
arrays with more that 200 emitters. 
A systematic study in the optical frequency detuning and coupling strength parameter space, 
shows that the region of chimera states lies between full synchronization and 
desynchronization. 
The ability to control the dynamics in and out of the synchronized state, may have
multiple technological applications
regarding the generation of on demand diverse waveforms~\cite{LIN04}. 
For future studies, it would be
worthwhile to explore this, as well as the effects introduced by
noise and the laser pump power, which is the most conveniently accessible
control parameter in chip scale diode systems.

\section*{Methods}
Recently, Kemeth \emph{et al.} presented a classification scheme for chimera states~\cite{kevrekidis}.
For measuring spatial coherence, in particular, they introduced a quantity called \emph{local curvature} which 
may be calculated at each time instance. This is done by applying the discrete Laplacian $DE$ on the spatial
data of the amplitude of the electric field:
\begin{equation}\label{eq9}
DE_{i}(t)=|E|_{i+1}(t)-2|E|_i(t)+|E|_{i-1}(t), \quad i=1\dots M. 
\end{equation}
In the synchronization regime the local curvature is close to zero while in the asynchronous
regime it is finite and fluctuating. Therefore, if $g$ 
is the normalized probability density function of $|DE|$, $g(|DE|=0)$ measures the relative size of spatially 
coherent regions in each temporal realization. For a fully synchronized system $g(|DE|=0)=1$,
while for a totally incoherent system it holds that $g(|DE|=0)=0$. A value between $0$ and $1$
of $g(|DE|=0)$ indicates coexistence of synchronous and asynchronous lasers.

The quantity $g$ is time-dependent. Complementary to the local curvature we also calculate the spatial extent occupied by the coherent lasers which is given by the following integral:
\begin{equation}\label{eq:g_0}
g_{0}(t)=\int_{0}^{\delta}g(t,|DE|)d|DE|,
\end{equation}
where $\delta$ is a threshold value distinguishing between coherence and incoherence
which is related to the maximum curvature and is system-dependent. We will apply these measures in order to classify the observed patterns and we will discuss their dependence on the coupling strength $H$ and the detuning parameter $\Delta$.

\section*{Author contributions statement}
J.~S., J.~H. and V.~K. conceived and designed 
the study and performed the numerical simulations. 
All authors carried out the analysis and wrote the article.

\section*{Acknowledgments}
This work was partially supported by
the European Union Seventh Framework Program (FP7-REGPOT-2012-2013-1)
under grant agreement no 316165,
the Ministry of Education and Science of the Russian Federation in the
framework of the Increase Competitiveness Program of NUST ``MISiS'' (No. K2-2015-007),
and the Ministry of Education and Science of the Republic of Kazakhstan via Contract number 339/76-2015.

\section*{Additional information}
The authors declare no competing financial interest. Correspondence or request for material should be addressed to J.~S., J.~H., V.~K. or  G.~P.~T. 
\

\begin{thebibliography}{10}
\expandafter\ifx\csname url\endcsname\relax
  \def\url#1{\texttt{#1}}\fi
\expandafter\ifx\csname urlprefix\endcsname\relax\def\urlprefix{URL }\fi
\providecommand{\bibinfo}[2]{#2}
\providecommand{\eprint}[2][]{\url{#2}}

\bibitem{Urbana}
\bibinfo{author}{Johnson, M.~T.}, \bibinfo{author}{Siriani, D.~F.},
  \bibinfo{author}{Tan, M.~P.} \& \bibinfo{author}{Choquette, K.~D.}
\newblock \bibinfo{title}{High-speed beam steering with phased vertical cavity
  laser arrays}.
\newblock \emph{\bibinfo{journal}{Journal of Selected Topics in Quantum
  Electronics}} \textbf{\bibinfo{volume}{19}}, \bibinfo{pages}{1701006}
  (\bibinfo{year}{2013}).

\bibitem{HEC13}
\bibinfo{author}{Heck, M.~J.}, et al.
\newblock \bibinfo{title}{Hybrid silicon photonic integrated circuit
  technology}.
\newblock \emph{\bibinfo{journal}{Journal of Selected Topics in Quantum
 Electronics}} \textbf{\bibinfo{volume}{19}}, \bibinfo{pages}{6100117}
 (\bibinfo{year}{2013}).


   
\bibitem{MIT_2016}
\bibinfo{author}{Kao, T.~Y.}, \bibinfo{author}{Reno, J.~L.} \&
  \bibinfo{author}{Hu, Q.}
\newblock \bibinfo{title}{Phase-locked laser arrays through global antenna
  mutual coupling}.
\newblock \emph{\bibinfo{journal}{Nature Photonics}}
  \textbf{\bibinfo{volume}{10}}, \bibinfo{pages}{1038} (\bibinfo{year}{2016}).

\bibitem{Fischer_2015}
\bibinfo{author}{Brunner, D.} \& \bibinfo{author}{Fischer, I.}
\newblock \bibinfo{title}{Recon gurable semiconductor laser networks based on
  diractive coupling}.
\newblock \emph{\bibinfo{journal}{Optics Letters}}
  \textbf{\bibinfo{volume}{40}}, \bibinfo{pages}{3854} (\bibinfo{year}{2015}).

\bibitem{kevrekidis}
\bibinfo{author}{Kemeth, F.~P.}, \bibinfo{author}{Haugland, S.~W.},
  \bibinfo{author}{Schmidt, L.}, \bibinfo{author}{Kevrekidis, I.~G.} \&
  \bibinfo{author}{Krischer, K.}
\newblock \bibinfo{title}{A classification scheme for chimera states}.
\newblock \emph{\bibinfo{journal}{Chaos}}
  \textbf{\bibinfo{volume}{26}}, \bibinfo{pages}{094815}  (\bibinfo{year}{2016}).

\bibitem{KUR02a}
\bibinfo{author}{Kuramoto, Y.} \& \bibinfo{author}{Battogtokh, D.}
\newblock \bibinfo{title}{Coexistence of coherence and incoherence in
  nonlocally coupled phase oscillators}.
\newblock \emph{\bibinfo{journal}{Nonlinear Phenomena in Complex Systems}}
  \textbf{\bibinfo{volume}{5}}, \bibinfo{pages}{380--385}
  (\bibinfo{year}{2002}).

\bibitem{panaggio:2015}
\bibinfo{author}{Pannagio, M.~J.} \& \bibinfo{author}{Abrams, D.}
\newblock \bibinfo{title}{Chimera states: coexistence of coherence and
  incoherence in networks of coupled oscillators}.
\newblock \emph{\bibinfo{journal}{Nonlinearity}} \textbf{\bibinfo{volume}{28}},
  \bibinfo{pages}{R67} (\bibinfo{year}{2015}).

\bibitem{tinsley:2012}
\bibinfo{author}{Tinsley, M.~R.}, \bibinfo{author}{Nkomo, S.} \&
  \bibinfo{author}{Showalter, K.}
\newblock \bibinfo{title}{Chimera and phase-cluster states in populations of
  coupled chemical oscillators}.
\newblock \emph{\bibinfo{journal}{Nature Physics}}
  \textbf{\bibinfo{volume}{8}}, \bibinfo{pages}{662--665}
  (\bibinfo{year}{2012}).

\bibitem{hagerstrom:2012}
\bibinfo{author}{Hagerstrom, A.~M.} 
\bibinfo{author}{Murphy, T.~E.}, \bibinfo{author}{Roy, R.}, \bibinfo{author}{H\"ovel, P}, \bibinfo{Omelchenko, I} \&
  \bibinfo{author}{Sch\"oll, E.}
\newblock \bibinfo{title}{Experimental observation of chimeras in coupled-map
  lattices}.
\newblock \emph{\bibinfo{journal}{Nature Physics}}
  \textbf{\bibinfo{volume}{8}}, \bibinfo{pages}{658--661}
  (\bibinfo{year}{2012}).

\bibitem{wickramasinghe:2013}
\bibinfo{author}{Wickramasinghe, M.} \& \bibinfo{author}{Kiss, I.~Z.}
\newblock \bibinfo{title}{Spatially organized dynamical states in chemical
  oscillator networks: Synchronization, dynamical differentiation, and chimera
  patterns}.
\newblock \emph{\bibinfo{journal}{PLoS ONE}} \textbf{\bibinfo{volume}{8}},
  \bibinfo{pages}{e80586} (\bibinfo{year}{2013}).

\bibitem{martens:2013}
\bibinfo{author}{Martens, E.~A.}, \bibinfo{author}{Thutupalli, S.},
  \bibinfo{author}{Fourri{\`e}re, A.} \& \bibinfo{author}{Hallatschek, O.}
\newblock \emph{\bibinfo{journal}{Proc. Nat. Acad. Sciences}}
  \textbf{\bibinfo{volume}{110}}, \bibinfo{pages}{10563}
  (\bibinfo{year}{2013}).

\bibitem{Rosin2014}
\bibinfo{author}{Rosin, D.~P.}, \bibinfo{author}{Rontani, D.},
  \bibinfo{author}{Haynes, N.~D.}, \bibinfo{author}{Sch\"oll, E.} \&
  \bibinfo{author}{Gauthier, D.~J.}
\newblock \bibinfo{title}{Transient scaling and resurgence of chimera states in
  networks of boolean phase oscillators}.
\newblock \emph{\bibinfo{journal}{Physical Review E}}
  \textbf{\bibinfo{volume}{90}}, \bibinfo{pages}{030902(R)}
  (\bibinfo{year}{2014}).

  \bibitem{schmidt:2014}
\bibinfo{author}{Schmidt, L.}, \bibinfo{author}{Sch\"onleber, K.},
  \bibinfo{author}{Krischer, K.} \& \bibinfo{author}{Garc\'ia-Morales, V.}
\newblock \bibinfo{title}{Coexistence of synchrony and incoherence in
  oscillatory media under nonlinear global coupling}.
\newblock \emph{\bibinfo{journal}{Chaos}} \textbf{\bibinfo{volume}{24}},
  \bibinfo{pages}{013102} (\bibinfo{year}{2014}).

\bibitem{Gambuzza2014}
\bibinfo{author}{Gambuzza, L.~V.}, et al.
\newblock \bibinfo{title}{Experimental investigation of chimera states with
  quiescent and synchronous domains in coupled electronic oscillators}.
\newblock \emph{\bibinfo{journal}{Physical Review E}}
  \textbf{\bibinfo{volume}{90}}, \bibinfo{pages}{032905}
  (\bibinfo{year}{2014}).

\bibitem{Kapitaniak2014}
\bibinfo{author}{Kapitaniak, T.}, \bibinfo{author}{Kuzma, P.},
  \bibinfo{author}{Wojewoda, J.}, \bibinfo{author}{Czolczynski, K.} \&
  \bibinfo{author}{Maistrenko, Y.}
\newblock \bibinfo{title}{Imperfect chimera states for coupled pendula}.
\newblock \emph{\bibinfo{journal}{Scientific Reports}}
  \textbf{\bibinfo{volume}{4}}, \bibinfo{pages}{6379} (\bibinfo{year}{2014}).

\bibitem{HAR16}
\bibinfo{author}{Hart, J.~D.}, \bibinfo{author}{Bansal, K.},
  \bibinfo{author}{Murphy, T.~E.} \& \bibinfo{author}{Roy, R.}
\newblock \bibinfo{title}{Experimental observation of chimera and cluster
  states in a minimal globally coupled network}.
\newblock \emph{\bibinfo{journal}{Chaos}} \textbf{\bibinfo{volume}{26}},
  \bibinfo{pages}{094801} (\bibinfo{year}{2016}).

\bibitem{LAI12}
\bibinfo{author}{Laing, C.~R.}, \bibinfo{author}{Rajendran, K.} \&
  \bibinfo{author}{Kevrekidis, I.~G.}
\newblock \bibinfo{title}{Chimeras in random non-complete networks of phase
  oscillators}.
\newblock \emph{\bibinfo{journal}{Chaos}} \textbf{\bibinfo{volume}{22}},
  \bibinfo{pages}{013132} (\bibinfo{year}{2012}).

\bibitem{YAO13}
\bibinfo{author}{Yao, N.}, \bibinfo{author}{Huang, Z.~G.},
  \bibinfo{author}{Lai, Y.~C.} \& \bibinfo{author}{Zheng, Z.~G.}
\newblock \bibinfo{title}{Robustness of chimera states in complex dynamical
  systems}.
\newblock \emph{\bibinfo{journal}{Scientific Reports}}
  \textbf{\bibinfo{volume}{3}}, \bibinfo{pages}{3522} (\bibinfo{year}{2013}).

\bibitem{OME15}
\bibinfo{author}{Omelchenko, I.}, \bibinfo{author}{Provata, A.},
  \bibinfo{author}{J.~Hizanidis, E.~S.} \& \bibinfo{author}{H\"ovel, P.}
\newblock \bibinfo{title}{Robustness of chimera states for coupled
  fitzhugh-nagumo oscillators}.
\newblock \emph{\bibinfo{journal}{Physical Review E}}
  \textbf{\bibinfo{volume}{91}}, \bibinfo{pages}{022917}
  (\bibinfo{year}{2015}).

\bibitem{HIZ16a}
\bibinfo{author}{Hizanidis, J.}, \bibinfo{author}{Kouvaris, N.~E.},
  \bibinfo{author}{Zamora-L\'opez, G.}, \bibinfo{author}{D\'iaz-Guilera, A.} \&
  \bibinfo{author}{Antonopoulos, C.~G.}
\newblock \bibinfo{title}{Chimera-like states in modular neural networks}.
\newblock \emph{\bibinfo{journal}{Scientific Reports}}
  \textbf{\bibinfo{volume}{6}}, \bibinfo{pages}{19845} (\bibinfo{year}{2016}).

\bibitem{SEM16}
\bibinfo{author}{Semenova, N.}, \bibinfo{author}{Zakharova, A.},
  \bibinfo{author}{Anishchenko, V.} \& \bibinfo{author}{Sch\"oll, E.}
\newblock \bibinfo{title}{Coherence-resonance chimeras in a network of
  excitable elements}.
\newblock \emph{\bibinfo{journal}{Physical Review Letters}}
  \textbf{\bibinfo{volume}{117}}, \bibinfo{pages}{014102}
  (\bibinfo{year}{2016}).

\bibitem{SIE14}
\bibinfo{author}{Sieber, J.}, \bibinfo{author}{Omelchenko, O.~E.} \&
  \bibinfo{author}{Wolfrum, M.}
\newblock \bibinfo{title}{Controlling unstable chaos: Stabilizing chimera
  states by feedback}.
\newblock \emph{\bibinfo{journal}{Physical Review Letters}}
  \textbf{\bibinfo{volume}{112}}, \bibinfo{pages}{054102}
  (\bibinfo{year}{2014}).

\bibitem{BIC15}
\bibinfo{author}{Bick, C.} \& \bibinfo{author}{Martens, E.~A.}
\newblock \bibinfo{title}{Controlling chimeras}.
\newblock \emph{\bibinfo{journal}{New Journal of Physics}}
  \textbf{\bibinfo{volume}{17}}, \bibinfo{pages}{033030}
  (\bibinfo{year}{2015}).

\bibitem{ISE15}
\bibinfo{author}{Isele, T.}, \bibinfo{author}{Hizanidis, J.},
  \bibinfo{author}{Provata, A.} \& \bibinfo{author}{H\"ovel, P.}
\newblock \bibinfo{title}{Controlling chimera states: The influence of
  excitable units}.
\newblock \emph{\bibinfo{journal}{Physical Review E}}
  \textbf{\bibinfo{volume}{93}}, \bibinfo{pages}{022217}
  (\bibinfo{year}{2016}).

\bibitem{OME16}
\bibinfo{author}{Omelchenko, I.}, \bibinfo{author}{Omelchenko, O.~E.},
  \bibinfo{author}{Zakharova, A.}, \bibinfo{author}{Wolfrum, M.} \&
  \bibinfo{author}{Sch{\"o}ll, E.}
\newblock \bibinfo{title}{Tweezers for chimeras in small networks}.
\newblock \emph{\bibinfo{journal}{Physical Review Letters}}
  \textbf{\bibinfo{volume}{116}}, \bibinfo{pages}{114101}
  (\bibinfo{year}{2016}).

\bibitem{mandel1}
\bibinfo{author}{Kozyreff, G.}, \bibinfo{author}{Vladimirov, A.} \&
  \bibinfo{author}{Mandel, P.}
\newblock \bibinfo{title}{Global coupling with time delay in an array of
  semiconductor lasers}.
\newblock \emph{\bibinfo{journal}{Physical Review Letters}}
  \textbf{\bibinfo{volume}{85}}, \bibinfo{pages}{3809} (\bibinfo{year}{2000}).

\bibitem{OLI01}
\bibinfo{author}{Oliva, R.~A.} \& \bibinfo{author}{Strogatz, S.~H.}
\newblock \bibinfo{title}{Dynamics of a large array of globally coupled lasers
  with distributed frequencies}.
\newblock \emph{\bibinfo{journal}{International Journal of Bifurcation and
  Chaos}} \textbf{\bibinfo{volume}{11}}, \bibinfo{pages}{2359}
  (\bibinfo{year}{2001}).

\bibitem{UCH01}
\bibinfo{author}{Uchida, A.}, \bibinfo{author}{Liu, Y.},
  \bibinfo{author}{Fischer, I.}, \bibinfo{author}{Davis, P.} \&
  \bibinfo{author}{Aida, T.}
\newblock \bibinfo{title}{Chaotic antiphase dynamics and synchronization in
  multimode semiconductor lasers}.
 \newblock \emph{\bibinfo{journal}{Physical Review A}}
  \textbf{\bibinfo{volume}{64}}, \bibinfo{pages}{023801} (\bibinfo{year}{2001}). 
   

\bibitem{DAH12}
\bibinfo{author}{Dahms, T.}, \bibinfo{author}{Lehnert, J.} \&
  \bibinfo{author}{Sch\"oll, E.}
\newblock \bibinfo{title}{Cluster and group synchronization in delay-coupled
  networks}.
\newblock \emph{\bibinfo{journal}{Physical Review E}}
  \textbf{\bibinfo{volume}{86}}, \bibinfo{pages}{016202}
  (\bibinfo{year}{2012}).

\bibitem{Fischer}
\bibinfo{author}{Soriano, M.~C.}, \bibinfo{author}{Garcia-Ojalvo, J.},
  \bibinfo{author}{Mirasso, C.~R.} \& \bibinfo{author}{Fischer, I.}
\newblock \bibinfo{title}{Complex photonics: Dynamics and applications of
  delay-coupled semiconductors lasers}.
\newblock \emph{\bibinfo{journal}{Reviews Modern Physics}}
  \textbf{\bibinfo{volume}{85}}, \bibinfo{pages}{421--470}
  (\bibinfo{year}{2013}).

\bibitem{LYT97}
\bibinfo{author}{Lythe, G.}, \bibinfo{author}{Erneux, T.},
  \bibinfo{author}{Gavrielides, A.} \& \bibinfo{author}{Kovanis, V.}
\newblock \bibinfo{title}{Low pump limit of the bifurcation to periodic
  intensities in a semiconductor laser subject to external optical feedback}.
\newblock \emph{\bibinfo{journal}{Physical Review A}}
  \textbf{\bibinfo{volume}{55}}, \bibinfo{pages}{4443--4448}
  (\bibinfo{year}{1997}).

\bibitem{PEC14}
\bibinfo{author}{Pecora, L.~M.}, \bibinfo{author}{Sorrentino, F.},
  \bibinfo{author}{Hagerstrom, A.~M.}, \bibinfo{author}{Murphy, T.~E.} \&
  \bibinfo{author}{Roy, R.}
\newblock \bibinfo{title}{Cluster synchronization and isolated
  desynchronization in complex networks with symmetries}.
\newblock \emph{\bibinfo{journal}{Nature Communications}}
  \textbf{\bibinfo{volume}{5}}, \bibinfo{pages}{4079} (\bibinfo{year}{2014}).

\bibitem{ALS96}
\bibinfo{author}{Alsing, P.~M.}, \bibinfo{author}{Kovanis, V.},
  \bibinfo{author}{Gavrielides, A.} \& \bibinfo{author}{Erneux, T.}
\newblock \bibinfo{title}{Lang and kobayashi phase equation}.
\newblock \emph{\bibinfo{journal}{Physical Review A}}
  \textbf{\bibinfo{volume}{53}}, \bibinfo{pages}{44294434}
  (\bibinfo{year}{1996}).

\bibitem{JUN16}
\bibinfo{author}{Junges, L.}, \bibinfo{author}{Gavrilides, A.} \&
  \bibinfo{author}{Gallas, J. A.~C.}
\newblock \bibinfo{title}{Synchronization properties of two mutually
  delay-coupled semiconductor lasers}.
\newblock \emph{\bibinfo{journal}{Journal of the Optical Society of America B}}
  \textbf{\bibinfo{volume}{33}}, \bibinfo{pages}{C65} (\bibinfo{year}{2016}).

\bibitem{LAR13}
\bibinfo{author}{Larger, L.}, \bibinfo{author}{Penkovsky, B.} \&
  \bibinfo{author}{Maistrenko, Y.}
\newblock \bibinfo{title}{Chimera states for delayed-feedback systems}.
\newblock \emph{\bibinfo{journal}{Physical Review Letters}}
  \textbf{\bibinfo{volume}{111}}, \bibinfo{pages}{054103}
  (\bibinfo{year}{2013}).

\bibitem{LAR15}
\bibinfo{author}{Larger, L.}, \bibinfo{author}{Penkovsky, B.} \&
  \bibinfo{author}{Maistrenko, Y.}
\newblock \bibinfo{title}{Laser chimeras as a paradigm for multistable patterns
  in complex systems}.
\newblock \emph{\bibinfo{journal}{Nature Communications}}
  \textbf{\bibinfo{volume}{6}}, \bibinfo{pages}{7752} (\bibinfo{year}{2015}).

\bibitem{BOE15}
\bibinfo{author}{B\"ohm, F.}, \bibinfo{author}{Zakharova, A.},
  \bibinfo{author}{Sch\"oll, E.} \& \bibinfo{author}{L\"udge, K.}
\newblock \bibinfo{title}{Amplitude-phase coupling drives chimera states in
  globally coupled laser networks}.
\newblock \emph{\bibinfo{journal}{Physical Review E}}
  \textbf{\bibinfo{volume}{91}}, \bibinfo{pages}{040901(R)}
  (\bibinfo{year}{2015}).

\bibitem{ROE16}
\bibinfo{author}{R\"ohm, A.}, \bibinfo{author}{B\"ohm, F.},  \&
  \bibinfo{author}{L\"udge, K.}
\newblock \bibinfo{title}{Small chimera states without multistability in a
  globally delay-coupled network of four lasers}.
\newblock \emph{\bibinfo{journal} {Physical Review E}} \textbf{\bibinfo{volume}{94}},
  \bibinfo{pages}{042204} (\bibinfo{year}{2016}).

\bibitem{HIZ16}
\bibinfo{author}{Hizanidis, J.}, \bibinfo{author}{Lazarides, N.} \&
  \bibinfo{author}{Tsironis, G.~P.}
\newblock \bibinfo{title}{Chimeras in locally coupled squids: Robust chimera states in SQUID metamaterials with local interactions}.
\newblock \emph{\bibinfo{journal}{Physical Review E}} \textbf{\bibinfo{volume}{94}},
  \bibinfo{pages}{032219} (\bibinfo{year}{2016}).

\bibitem{Dutta}
\bibinfo{author}{Blackbeard, N.}, \bibinfo{author}{Wieczoreka, S.},
  \bibinfo{author}{Erzgr\"aber, H.} \& \bibinfo{author}{Dutta, P.~S.}
\newblock \bibinfo{title}{From synchronisation to persistent optical turbulence
  in laser arrays}.
\newblock \emph{\bibinfo{journal}{Physica D}} \textbf{\bibinfo{volume}{43}},
  \bibinfo{pages}{286--287} (\bibinfo{year}{2014}).

\bibitem{Arecchi}
\bibinfo{author}{Arecchi, F.}, \bibinfo{author}{Lippi, G.},
  \bibinfo{author}{Puccioni, G.} \& \bibinfo{author}{Tredicce, J.}
\newblock \bibinfo{title}{Deterministic chaos in laser with injected signal}.
\newblock \emph{\bibinfo{journal}{Optics Communications}}
  \textbf{\bibinfo{volume}{51}}, \bibinfo{pages}{308--314}
  (\bibinfo{year}{1984}).

\bibitem{Wieczorek}
\bibinfo{author}{Wieczorek, S.}, \bibinfo{author}{Krauskopf, B.},
  \bibinfo{author}{Simpson, T.~B.} \& \bibinfo{author}{Lenstra, D.}
\newblock \bibinfo{title}{The dynamical complexity of optically injected
  semiconductor lasers}.
\newblock \emph{\bibinfo{journal}{Physics Reports}}
  \textbf{\bibinfo{volume}{416}}, \bibinfo{pages}{1--128}
  (\bibinfo{year}{2005}).

\bibitem{Simpson}
\bibinfo{author}{Simpson, T.~B.}, \bibinfo{author}{Liu, J.~M.},
  \bibinfo{author}{Gavrielides, A.}, \bibinfo{author}{Kovanis, V.} \&
  \bibinfo{author}{Alsing, P.~M.}
\newblock \bibinfo{title}{Period-doubling cascades and chaos in a semiconductor
  laser with optical injection}.
\newblock \emph{\bibinfo{journal}{Physical Review A}}
  \textbf{\bibinfo{volume}{51}}, \bibinfo{pages}{4181} (\bibinfo{year}{1995}).

\bibitem{Winful_1988}
\bibinfo{author}{Winful, H.~G.} \& \bibinfo{author}{Wang, S.~S.}
\newblock \bibinfo{title}{Stability of phase locking in coupled semiconductor
  laser arrays}.
\newblock \emph{\bibinfo{journal}{Applied Physics Letters}}
  \textbf{\bibinfo{volume}{53}}, \bibinfo{pages}{1894} (\bibinfo{year}{1988}).

\bibitem{Erneux_1997}
\bibinfo{author}{Kuske, R.} \& \bibinfo{author}{Erneux, T.}
\newblock \bibinfo{title}{Localized synchronization of two coupled solid state
  lasers}.
\newblock \emph{\bibinfo{journal}{Optics Communications}}
  \textbf{\bibinfo{volume}{139}}, \bibinfo{pages}{125--131}
  (\bibinfo{year}{1997}).

\bibitem{Roy_1997}
\bibinfo{author}{Thornburg, K.~S.}, \bibinfo{author}{Moeller, M.} \&
  \bibinfo{author}{Roy, R.}
\newblock \bibinfo{title}{Chaos and coherence in coupled lasers}.
\newblock \emph{\bibinfo{journal}{Physical Review E}}
  \textbf{\bibinfo{volume}{55}}, \bibinfo{pages}{3865} (\bibinfo{year}{1997}).

\bibitem{Roy_2007}
\bibinfo{author}{Rogister, F.} \& \bibinfo{author}{Roy, R.}
\newblock \bibinfo{title}{Localized excitations in arrays of synchronized laser
  oscillators}.
\newblock \emph{\bibinfo{journal}{Physical Review Letters}}
  \textbf{\bibinfo{volume}{98}}, \bibinfo{pages}{104101}
  (\bibinfo{year}{2007}).

\bibitem{Winful_1992}
\bibinfo{author}{Winful, H.~G.}
\newblock \bibinfo{title}{Instability threshold for an array of coupled
  semiconductor lasers}.
\newblock \emph{\bibinfo{journal}{Physical Review A}}
  \textbf{\bibinfo{volume}{46}}, \bibinfo{pages}{6093} (\bibinfo{year}{1992}).

\bibitem{ZAK14}
\bibinfo{author}{Zakharova, A.}, \bibinfo{author}{Kapeller, M.} \&
  \bibinfo{author}{Sch\"oll, E.}
\newblock \bibinfo{title}{Chimera death: Symmetry breaking in dynamical
  networks}.
\newblock \emph{\bibinfo{journal}{Physical Review Letters}}
  \textbf{\bibinfo{volume}{112}}, \bibinfo{pages}{154101}
  (\bibinfo{year}{2014}).

\bibitem{LIN04}
\bibinfo{author}{Lin, F.-Y.} \& \bibinfo{author}{Liu, J.-M.}
\newblock \bibinfo{title}{Diverse waveform generation using semiconductor
  lasers for radar and microwave applications}.
\newblock \emph{\bibinfo{journal}{IEEE Journal of Quantum Electronics}}
  \textbf{\bibinfo{volume}{40}}, \bibinfo{pages}{682} (\bibinfo{year}{2004}).

\end{thebibliography}
\end{document}